\documentclass[12pt]{article}
\usepackage{epsfig}

\newcommand{\rf}[1]{(\ref{#1})}
\newcommand{\eq}{\begin{equation}}
\newcommand{\eqx}{\end{equation}}
\newcommand{\eqn}{\begin{eqnarray}}
\newcommand{\eqnx}{\end{eqnarray}}
\newcommand{\f}[2]{\frac{#1}{#2}}

\newcommand{\tr}{\mbox{\rm tr}\,}

\newcommand{\al}{\alpha}

\newcommand{\Dl}{\Delta}
\newcommand{\dl}{\delta}

\newcommand{\Gm}{\Gamma}
\newcommand{\lm}{\lambda}
\newcommand{\kap}{\kappa}

\newcommand{\NN}{{\cal N}}

\newcommand{\hyp}[3]{{}_1 F_1 \left({#1},{#2};{#3}\right)}

\newcommand{\la}{\left\langle}
\newcommand{\ra}{\right\rangle}

\begin{document}


\hfill  December 2003

\begin{center}
\vspace{24pt}
{ \large \bf The decay of quantum D-branes}

\vspace{30pt}

{\sl Jan Ambj\o rn}$\,^{a,}$\footnote{Permanent address: 
The Niels Bohr Institute, Blegdamsvej 17, DK-2100 Copenhagen \O , Denmark.
{email: ambjorn@nbi.dk}.} and 
{\sl Romuald A. Janik}$\,^{b,}$\footnote{Permanent address: 
Institute of Physics, Jagellonian University,
Reymonta 4, PL 30-059 Krakow, Poland.
{email: ufrjanik@if.uj.edu.pl}  }

\vspace{24pt}
{\small

$^a$~The Perimeter Institute, \\
35 King Street North,\\
Waterloo, ON, N2J 2W9,  Canada\\
{\it email: ambjorn@nbi.dk}

\vspace{10pt}
$^b$~The Niels Bohr Institute, \\
Blegdamsvej 17, \\
DK-2100 Copenhagen \O , Denmark\\
{\it email: janik@nbi.dk}
}
\vspace{48pt}

\end{center}


\begin{center}
{\bf Abstract}
\end{center}
We study the {\em quantum} decay of D0-branes in two-dimensional 
0B string theory. The quantum nature of the branes provides
a natural cut-off for the closed string emission rate. We find exact
quantum mechanical wavefunctions for the decaying branes and show how
one can include the effects of the Fermi sea for any string coupling
(Fermi energy). 

\vspace{12pt}
\noindent


\newpage


\section{Introduction}

Originally dynamical triangulations of string 
world-sheets were introduced as a reparametrization
invariant lattice regularization. It successfully  defined
non-critical strings of fixed world-sheet topology
in a non-perturbative way,  the scaling to the continuum controlled
by the lattice spacing \cite{adf,david,kkm}.
The so-called matrix model description was a very convenient 
way of implementing the combinatorial task of summing over 
all abstract triangulations
of the world-sheet, which automatically, via the large $N$ expansion,
arranged the world-sheets according to topology.   
No physical interpretation of the matrix itself was given
until recently when it was suggested  in the case of the $c=1$
matrix model that the matrix  could be given the 
interpretation as the open string tachyon field between $N$ unstable  
D-branes ($N$ being the size of the matrix), the unstable D-branes
themselves being identified as the eigenvalues of the matrices 
\cite{verlinde,kms}.

This intriguing picture has passed a number of non-trivial tests
(also for models with $c < 1$, \cite{martinec,akk,km2}) 
and it offers the possibility 
for the first time to study {\it quantum D-branes} in {\it strongly} coupled
string theories.

In this paper we will discuss the decay of such quantum D-branes.

The outline of this paper is as follows. In section 2 we describe the
decay of the D0-brane in the classical approximation and reproduce
directly from the classical motion the closed string tachyon 1-point
function. In section~3 we present the exact quantum mechanical
treatment of the same process, although without taking into account
the effect of the Fermi sea, and show how a natural cut-off arises for
the closed string emission. In section 4 we show how to exactly
incorporate the effects of the string interactions (Fermi sea) in the
preceeding quantum mechanical picture. We close the paper with a
discussion. 

\section{Classical decay of the D0-brane}

In the double scaling limit the ground state of 2d {\em bosonic} string theory 
is constructed by filling one side of the upside-down harmonic oscillator 
potential $-\lm^2/2\al'$ with free fermions up to a Fermi level $-\mu$,
where zero energy is the top of the potential. The string coupling 
$g_s \sim 1/\mu$ and the partition function, closed string 
tachyon operators, macroscopic loop operators etc have
a unique perturbative expansion in $g_s$ which can be obtained
from the exactly solvable quantum mechanics of the upside-down harmonic 
oscillator. As first pointed out in \cite{ajk} the situation
is unclear if we move to strong string couplings, i.e. small $\mu$
near the top of the potential. Clearly tunneling 
(so-called ``non-perturbative'' effects) between the two sides is no longer
exponentially suppressed, but there is no ``first  principle'' 
in the bosonic case telling us how to relate the two sides. 

Fortunately in the case of two-dimensional 0B and 0A superstrings, as
realized in \cite{tt,many}, this ambiguity is lifted and both sides
are filled up to the same level.

In the following we will perform calculations within the 
0B model\footnote{However, all results derived here away from the 
strong coupling region ($g_s$ large) are qualitatively correct also 
for the the 2d bosonic string (the $c=1$ matrix model).} defined by
\eq
\int dT \, e^{-\int dt \, \left\{ \f{1}{2} (DT)^2 +V(T)\right\}}
\eqx
where the nondynamical gauge field in the covariant derivative just
restricts the path integral to singlet states and so the standard free
fermions give a complete description of all the degrees of freedom.

Further, the quantum states of  a D0-brane are precisely the quantum 
states of the Hamiltonian of the inverted harmonic potential except
that the spectrum starts at $-\mu$ and the model provides us 
with a complete description of the dynamics of a single D0-brane
\cite{sen1}.

The potential for the 0B matrix model differs from the pure 
bosonic potential by a factor of two:
\eq
V=-\f{1}{4\al'} \lm^2
\eqx
As realized in \cite{verlinde} eigenvalues may be identified with
(unstable) D0-branes. Consequently time dependent solutions are of
interest since they may be identified with decay processes.
A solution of the classical
equations of motion is
\eq
\label{e.eigclas}
\lm(t)=\sqrt{4\mu \al'} \sin \pi \tilde{\lm}  \cosh \f{t}{\sqrt{2\al'}},
\eqx
where $\tilde{\lm}$ should not be confused with the eigenvalue $\lm$.
Its energy relative to the Fermi surface 
$E_F=-\mu$ is $\mu \cos^2 \pi \tilde{\lm}$.
This solution is the matrix model analog of Sen's  
`rolling tachyon' \cite{sen} and it will emit closed strings 
tachyons, the rate dictated by its contribution to the on-shell 
closed string 1-point function. This in turn can be computed since the
{\em closed} string tachyon (in the NS-NS sector) 
can be defined \cite{tt} in the 0B matrix model as\footnote{In \rf{tachyon}
we have made a rotation $l \to il$, in accordance with the 
discussion in \cite{moore}, sec.~11.} 
\eq\label{tachyon}
T_{NS-NS}(E) \sim \lim_{l\to 0} \, (\mbox{\rm leg factor}) \cdot \int dt
\, e^{iE t} \, \la \tr e^{-i l \Phi^2(t)}\ra
\eqx
where  the so-called leg-factor is 
\eq
\label{e.legf}
\mbox{\rm leg factor} = \f{(2il)^{iE\sqrt{\al'/2}}}{\Gm(-iE\sqrt{\al'/2})},
\eqx
and the expectation value is with respect to the (double scaling limit)
matrix integral:
\eq\label{jan1}
\la \tr e^{-i l \Phi^2(t)}\ra = \int d\lm \; 
\rho_{ds}(\lm,t) \, e^{-i l \lm^2},
\eqx
$\rho_{ds}(\lm,t)$ being the appropriate double scaling limit of 
the eigenvalues.

Let us compute the contribution of the classical motion $\lm(t)$ of the
eigenvalue (\ref{e.eigclas}) to the 1-point function of the
NS-NS closed string tachyon. The contribution of such a 
classical eigenvalue to $\rho_{ds}(\lm,t)$ is simply $\dl (\lm-\lm(t))$. 
Thus we  have to evaluate the integral
\eq
\int dt\, e^{iE t} \, e^{-il 4\mu \al' \,\sin^2 \pi \tilde{\lm} \, \cosh^2
  \f{t}{\sqrt{2\al'}}} 
\eqx
in the limit $l \to 0$. The non-trivial small $l$ behavior comes 
from the the large $t$ region of the integral and the following 
change of integration variable 
\eq\label{var-u}
u= l e^{\sqrt{\f{2}{\al'}} t}
\eqx
leads to an integral
\eq
\sqrt{\f{\al'}{2}} l^{-iE  \sqrt{\f{\al'}{2}}} \int_0^\infty \f{du}{u} e^{-i
\mu \al' u\sin^2 \pi \tilde{\lm}} u^{iE \sqrt{\f{\al'}{2}}}.
\eqx
Performing the integral, setting $\al'=2$ and inserting the 
leg-factor (\ref{e.legf}) we obtain the correct answer:
\eq
\label{e.clasres}
e^{-iE \log \sin^2 \pi \tilde{\lm}} \f{\Gm(iE)}{\Gm(-iE)} \mu^{-iE}.
\eqx

Analogous calculations here and below, can of course be easily done
also for the closed string tachyons in the R-R sector.

\section{`Quantum' decay - D-brane wave packets}

As already mentioned 
the quantum theory of a D0-brane in the $c=1$ theory is completely 
described as a  free fermion with the Hamiltonian
\eq
\label{e.ham}
H=-\f{1}{2}\f{d^2}{d\lm^2} -\f{1}{2} \kap^2 \lm^2
\eqx 
where $\kap=1/\sqrt{2\al'}$. 
Therefore the classical picture of D0-brane decay is 
only approximate, limited by the uncertainty principle. 
In the quantum theory we  
are forced to consider wave packets instead of localized
eigenvalues following a classical trajectory with both a definite position and
momentum. 
The only non-trivial 
aspect is the fermionic nature of the D0-branes, which gives an indirect 
interaction with the Fermi-sea background and thus reflects the
nonzero value of the string coupling.

Let $\psi(\lm,t)$ be a normalized 
solution to the Schr\"{o}dinger equation for $H$
and let us assume it has no overlap with the eigenfunctions of $H$
with energy $E < E_F$. The contribution of this quantum state to the 
density $\rho_{ds}(\lm,t)$ will be $|\psi(\lm,t)|^2$ and thus the  
corresponding contribution to \rf{jan1}, when inserted in   
\rf{tachyon}, leads to the following integral:
\eq\label{Q}
T_{NS-NS}^{quantum}(E)=\lim_{l\to 0} \,\, (\mbox{\rm leg factor}) \cdot 
\int dt\, e^{iE t} \, \int d\lm e^{-il \lm^2} |\psi(\lm,t)|^2.
\eqx
For an arbitrary solution to the Schr\"{o}dinger equation 
we can always write
\eq\label{jan2}
\psi(\lm,t) = \sum_E c_E \psi_E(\lm,t).
\eqx
If this expansion of $\psi$ contains $E < E_F$ they have 
to be cut away and the wave function re-normalized (see section 4 below).

Let us concentrate on the simplest wave packets, Gaussian wave packets,
and ignore at first the problem of overlap with the Fermi sea.
We will address it in the next section. 
First a few general observations:
(1) for a quadratic potential the expectation
values $\la \lm(t) \ra$ will always follow a  classical orbit as 
follows from Ehrenfest's theorem. (2) A wave packet which is Gaussian 
at some time $t$ will remain Gaussian. This follows because the 
propagator $G(\lm,\lm';t)$ corresponding to $H$ is Gaussian in $\lm,\lm'$.
(3) Since the peak will then coincide with the expectation value of 
$\lm$ we know that for an  initial wave packet of the form 
\eq\label{e.init}
\psi(\lm,0)=\left( \f{2a}{\pi} \right)^{\f{1}{4}} e^{-a(\lm-\lm_0)^2+ip_0\lm}
\eqx
we have
\eq\label{jan3}
|\psi(\lm,t)|^2= \left( \f{2a(t)}{\pi} \right)^{\f{1}{2}} 
e^{-2a(t)(\lm-\lm(t))^2}
\eqx
 where $\lm(t)$ is just the classical orbit corresponding to 
initial values of $\lm_0,p_0$ and
$a(t)$ can be calculated to be\footnote{It can easily be read off from
the propagator 
\eq\label{prop}
G(\lm,\lm';t)=\left[\f{\kap}{2i\pi \sinh \kap t}\right]^{1/2}\;
\exp\left\{ \f{i\kap}{\sinh \kap t} [ (\lm^2+{\lm'}^2) \cosh \kap t -
2\lm\lm']\right\},
\eqx
\eq\label{psi-t}
\psi(\lm,t)= \int d\lm' \; G(\lm,\lm';t)\psi(\lm',0).
\eqx} 
\eq\label{at}
a(t) = \f{a}{\Dl(t)},~~~~
\Dl(t)=\cosh^2 \kap t +\f{4 a^2}{\kap^2} \sinh^2 \kap t
\eqx
Consequently the temporal evolution of wave packets is 
essentially dictated by the classical energy\footnote{The actual quantum
energy of the wave packet is $\la \psi|H|\psi\ra = 
H_{cl}(\lm_0,p_0)+\f{a}{2}-\f{\kap^2}{8a} $.} $H_{cl}(\lm_0,p_0)= 
\f{1}{2}(p_0^2-\kap^2 \lm_0^2)$ since the peak just follows the classical 
orbit and the wave packet never splits in two, as a generic wave
packet would do in the inverted harmonic potential. However, 
due to the very rapid spread of the wave one still has non-zero
transmission though  the potential barrier even if $H_{cl} < 0$.
For a given initial wave packet we see 
that a  true classical picture of a localized wave function is only 
valid for times times $\kap t < \log a$ or $t < \sqrt{\al'}\log a$.

Let us for simplicity consider a wave packet where $p_0 =0$ and $\lm_0 >0$.
This is the wave packet version of Sen's rolling tachyon and {\it $\lm_0$ will
now be related to $\mu$ as follows}:
\eq\label{x-mu}
\lm_0^2= 4\mu\al' \sin^2 \pi \tilde{\lm}.
\eqx 
We can now calculate the closed string tachyon one-point function \rf{Q}.
The integral over the eigenvalues gives in this case
\eq
\f{1}{\sqrt{1+i
    \f{\Dl}{2a} l}}\;
\exp{\Big(-il \f{\lm_0^2 \cosh^2 \kap t}{1+i \f{\Dl}{2a} l}\Big)}
 \eqx
It might seem as if one could neglect the contribution 
$ l \Dl/2a$ from the spreading of the wave packet
since we take the $l \to 0$ limit. However since $\Dl$ grows
exponentially in time this contribution is of the same order as the
classical piece. Therefore the classical result (\ref{e.clasres}) will
get modified. Making the change of variable \rf{var-u}
we have to perform now the integral
\eq
\sqrt{\al'/2} l^{-iE\sqrt{\al'/2}} \int_0^\infty \f{du}{u} \;
u^{iE \sqrt{\al'/2}} \cdot
\f{e^{-i\f{\lm_0^2}{4} \f{u}{1+iC u}}}{\sqrt{1+iCu}}
\eqx
where
\eq\label{C}
C=\f{1}{4} \left(\f{1}{2a}+ \f{2a}{\kap^2} \right)= 
\f{1}{2} \sqrt{\al'/2} \left(\f{1}{4a\sqrt{\al'/2}}+ 
{4a}\sqrt{\al'/2} \right) .
\eqx 
This integral can be performed by a further change  
of variable $v=\f{iC u}{1+iCu}$ which leads to an integral of the form
(with  $\al'=2$)
\eq
(iC)^{-iE} \int_0^1 dv v^{iE-1} (1-v)^{-iE-\f{1}{2}} e^{-\f{\lm_0^2}{4C}.
  v}
\eqx
The final result, after including the leg-factors, is  
\eqn
quantum &=& \f{\Gm(i E)}{\Gm(-iE)} \cdot \left(\f{C}{2}\right)^{-iE}
\Gm\left( \f{1}{2}-iE \right) \f{1}{\sqrt{\pi}}
\hyp{iE}{\f{1}{2}}{-\f{\lm_0^2}{4C}}\\ 
classical &=& \f{\Gm(i E)}{\Gm(-iE)} \cdot \left( \f{\lm_0^2}{8}
\right)^{-iE} 
\eqnx
where for comparison we have also written the classical result and where 
the relation between $\lm_0$ and $\mu$ is as in \rf{x-mu}.
Note that this appearance of $\mu$ (equivalently $g_s$) here is purely
`kinematical' and enters only through the initial condition
$\lm_0$. The appearance of $\mu$ due to string interactions (influence
of Fermi sea) will be treated in the next section.

Let us first discuss the case when $\lm_0\gg 0$.

First note that from the asymptotics
$\hyp{a}{b}{-x} \sim x^{-a} \Gm(b)/\Gm(b-a)$ we see that  the quantum result
for the tachyon 1-point function leads to the classical one in the
limit $\mu \to \infty$ ($g_s \to 0$) while keeping the energies $E$
fixed.

However for any large but finite $\mu$ the behavior of both formulas
is quite different. The classical result is a pure phase
factor which for large $E$ behaves as:
\eq\label{cl-osc}
T_{NS-NS}^{classical}(E)\sim -i (\ldots)^{-iE} \cdot e^{-2iE} e^{2i E
  \log E} +{\cal O}(1/E) 
\eqx  
Thus the number of emitted particles, 
$N = \int \f{dE}{E} | T_{NS-NS}(E)|^2 $, diverges logarithmically 
and the expectation value of the emitted energy diverges linearly
\cite{kms} as in the similar calculation in 26 dimensions \cite{llm}.
It was suggested that a cut-off of order $1/g_s$ should be put into
these calculations but  
in our  case of quantum D0-branes we see that such a cut-off arises naturally. 
Indeed, denoting $M=\lm_0^2/(4C)$ we have the large $E$ asymptotics
\eqn
\hyp{iE}{\f{1}{2}}{-M} &\sim& e^{-\f{M}{2}} \cosh\left( 2\sqrt{M}
  \sqrt{\f{1}{4}-iE} \right) \\
\label{e.gamasymp}
\Gm(1/2-iE) &\sim& \sqrt{2\pi} e^{-\f{\pi}{2}E} e^{iE -iE\log E}
\eqnx 
We see that the answer is cut off when $\pi E/2 \gg
\sqrt{2ME}$, or, dropping constants,  when $E \gg M$.
{\it Thus the energy emitted is finite, the regularization 
provided by the quantum mechanical nature of the D0-brane}
as indeed conjectured in \cite{kms}.
If the classical orbit has a turning point away from zero
($\lm_0\gg 0$) then the cut-off is (with $\al'$ reinserted 
and $\mu \sqrt{\al'} \sim 1/g_s$)
\eq\label{cutoff}
\sqrt{\al'} E_{cutoff} \sim \f{1}{g_s} \; 
\f{1}{c a \sqrt{\al'}+({1}/{c a\sqrt{\al'}})} \sim \f{1}{g_s},
\eqx
where the constant $c$ is of order 1 and the last $\sim$ is valid when 
the location of the wave packet at $t=0$ is of the order of $\sqrt{\al'}$.
Anyway, $1/g_s$ will always serve as a upper cut-off as long as the 
$\lm_0 \gg 0$.
 
Let us now turn to the opposite case.
When the turning point of the classical orbit is close to the 
maximum of the potential $M$ goes to zero and the cut-off is now set
just by (\ref{e.gamasymp}). Then in the r.h.s. of (\ref{cutoff}) 
$1/g_s$ should be replaced by 1,
showing that quantum effects wash out the signature of 
string perturbation expansion.
In particular this is the case in the ``pure quantum case''
where $\lm_0=0$ and the classical eigenvalue is located
on the top of the potential without rolling down. This situation
is not sustainable for the quantum brane and the probability 
of finding the eigenvalue within a distance $d$ from the origin
decreases like $e^{-\kap t}d\sqrt{a}$ for the Gaussian wave packet.
The cut-off of the emitted energy is then just $E_{cutoff} = \sqrt{1/\al'}$.

\section{Large $g_s$ and  inclusion of the Fermi sea}

Above we obtained the exact motion of the wave packet in the
inverted potential. However we treated the motion independently from
the other eigenvalues. This is a good approximation if the overlap with
the Fermi sea is small. If we consider wave packets imitating to some degree
the rolling tachyon of Sen we expect the above calculation to be 
reliable if $\mu\sqrt{\al'} \sim 1/g_s$ is large. 
However, when the string coupling 
is large the Fermi level is close to the top and we are bound 
to get a significant overlap between the wave packet and the Fermi 
sea and it has to be taken into account. 
     
The {\em exact} $N$-body wave function of the
system will just be a Slater determinant with the $\psi(\lm,t)$
wave function derived in the previous section as one of its components.
Then any observable such as the closed string tachyon 1-point function has to
be computed using the Slater determinant wave function. 
Such a calculation would be in general quite formidable due to the
needed anti-symmetrization and lots of possible subtle cancellations.
This will introduce another source for the dependence on the string coupling
$g_s$ into the picture (since this is encoded in the Fermi level). 

In order to bypass these complications we will construct from
$\psi(\lm,t)$ an equivalent $\mu$-dependent wave function
$\psi_{eff}(\lm,t)$ which will render anti-symmetrization trivial and
allow to use single particle intuitions.

To this end let us consider properly normalized Slater determinants. If
all the component wave functions are orthogonal to each other then the
Slater determinant will be properly normalized. This is certainly the
case for the levels in the Fermi sea, but the wave packet $\psi(\lm,t)$
will have components also below the Fermi level. Of course these
components will not contribute to the $N$-body wave function. So the
only effect of the Fermi sea will be to truncate the original
expansion of the wave packet
\eq
\label{e.decomp}
\psi(\lm,t) =\sum_E c_E \;e^{-iEt} \phi_E(\lm)
\eqx
to
\eq
\psi_{eff}(\lm,t) = \NN \cdot \psi_{proj}(\lm,t) =\NN \sum_{E>-\mu} c_E
\;e^{-iEt} \phi_E(\lm) 
\eqx
where the normalization constant is
\eq\label{NN}
\NN = \left( \sum_{E>-\mu} |c_E|^2 \right)^{-\f{1}{2}}
\eqx 

Note that the normalization constant $\NN$ will re-normalize all
expectation values (average energies, 1-point functions) from the
single particle case. The energy profiles of the 1-point functions
will of course also change due to the absence in $\psi_{proj}(\lm,t)$ of
some of the original components of $\psi(\lm,t)$.
Since $\psi_{eff}(\lm,t)$ is orthogonal to all
states below the Fermi level anti-symmetrization is trivial and
effectively drops out.

The projector is $P_{E_F}(E)= \theta (E-E_F)$, the Fourier transform
of which is
\eq
P_{E_F}(t,s) = \f{1}{2} \delta(t-s) + \f{i e^{iE_F (t-s)}}{2\pi (t-s)}. 
\eqx
Thus we can write (with $E_F = -\mu$) 
\eq\label{int-rep}
\psi_{proj}(\lm,t) =\f{1}{2} \psi(\lm,t) +\f{i}{2\pi} \int_0^\infty
\f{ds}{s} \left[ e^{-i\mu s} \psi(\lm,t+s) - e^{i\mu s} \psi(\lm,t-s)
\right].
\eqx
This integral representation is of course completely general
and might be convenient whenever
one actually knows the wave function $\psi(\lm,t)$.

\begin{figure}
\vspace{-2.5cm}
\centerline{%
\hspace{-2cm}
\epsfysize=6cm \epsfbox{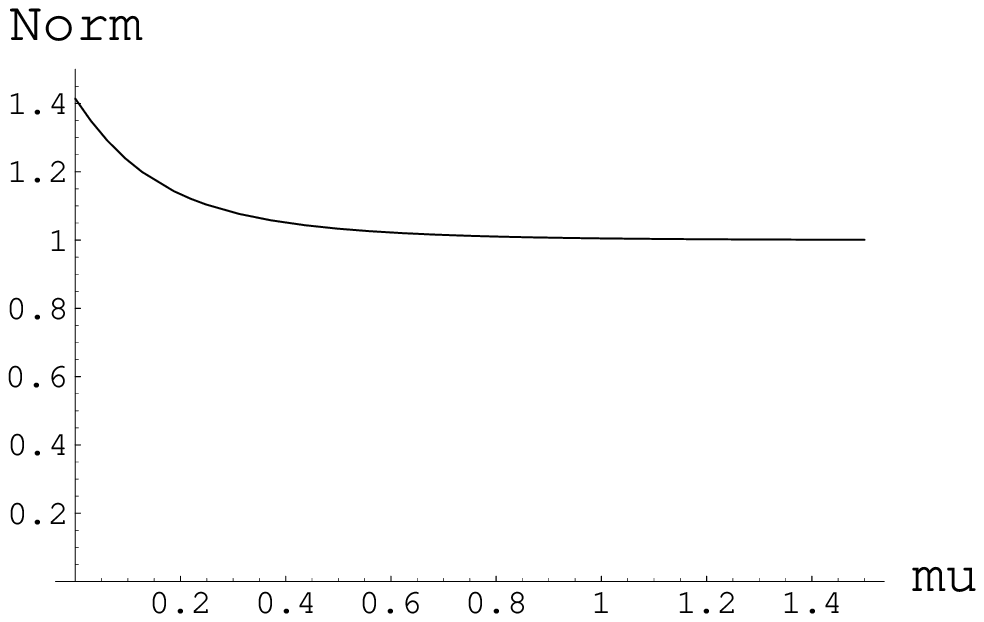}\hfill
\epsfysize=6cm \epsfbox{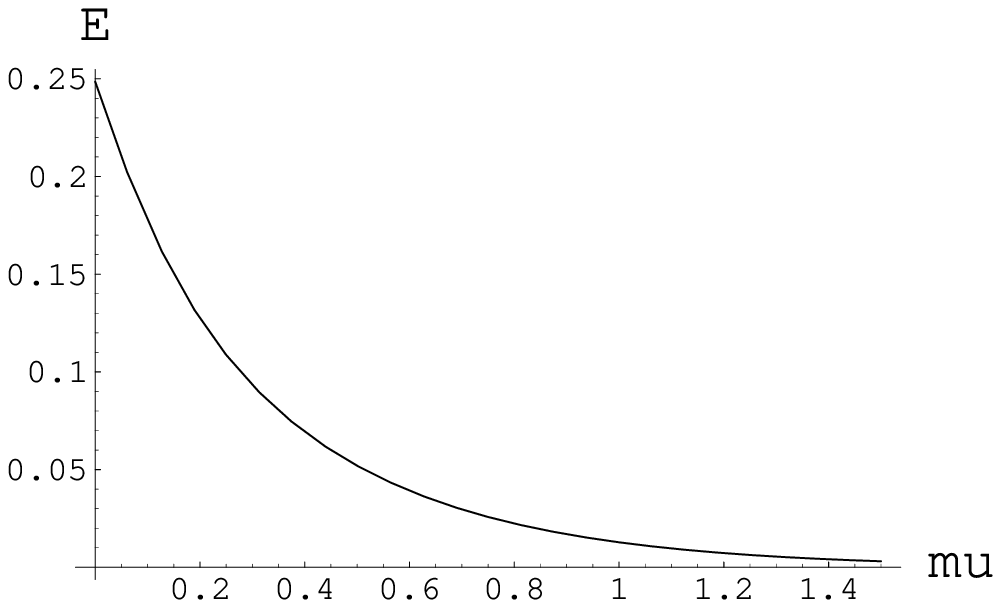}\hfill}
\caption{The normalization constant ${\cal N}$ and the energy of the
  effective wave-function $\psi_{eff}(\lm,t)$ (corresponding to the 
Gaussian wave packet \rf{psi0}) as a function of $\mu$.} 
\label{fig1}
\end{figure}

The simplest case, and the one in the family of states
we have considered here which is {\it least} semi-classical, is 
\eq\label{psi0}
\psi_0(\lm,0)= \f{1}{(2\pi)^{\f{1}{4}}} e^{-\f{\lm^2}{4}},~~~~~~
\psi_0(\lm,t)=\f{1}{(2\pi)^{\f{1}{4}}} \f{1}{\sqrt{\cosh \f{t}{2}}}
\f{e^{-\f{\lm^2}{4} \f{1-i \tanh \f{t}{2}}{1+i \tanh \f{t}{2}}}}{
  \sqrt{1+i \tanh \f{t}{2}}}, 
\eqx
where the last expression follows from \rf{psi-t}. 
With this choice of wave function both 
the classical energy $H_{cl}=\f{1}{2}(p_0^2-\kap^2 \lm_0^2)$
and the quantum energy $\la \psi_0|H|\psi_0\ra = 
H_{cl}(\lm_0,p_0)+\f{a}{2}-\f{\kap^2}{8a} $ are zero. 

\begin{figure}
\vspace{-2.5cm}
\centerline{%
\hspace{-2cm}
\epsfysize=6cm \epsfbox{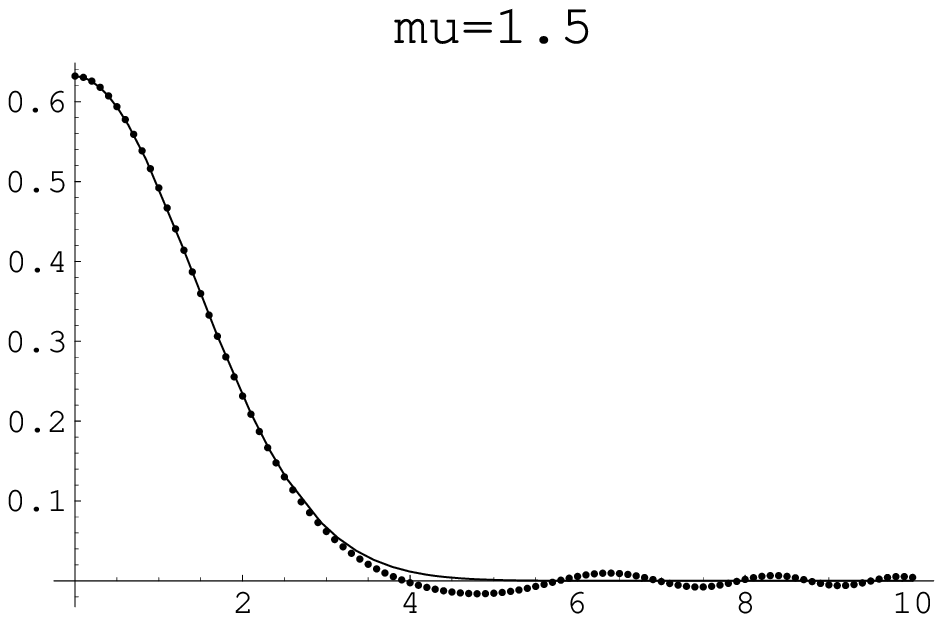}\hfill
\epsfysize=6cm \epsfbox{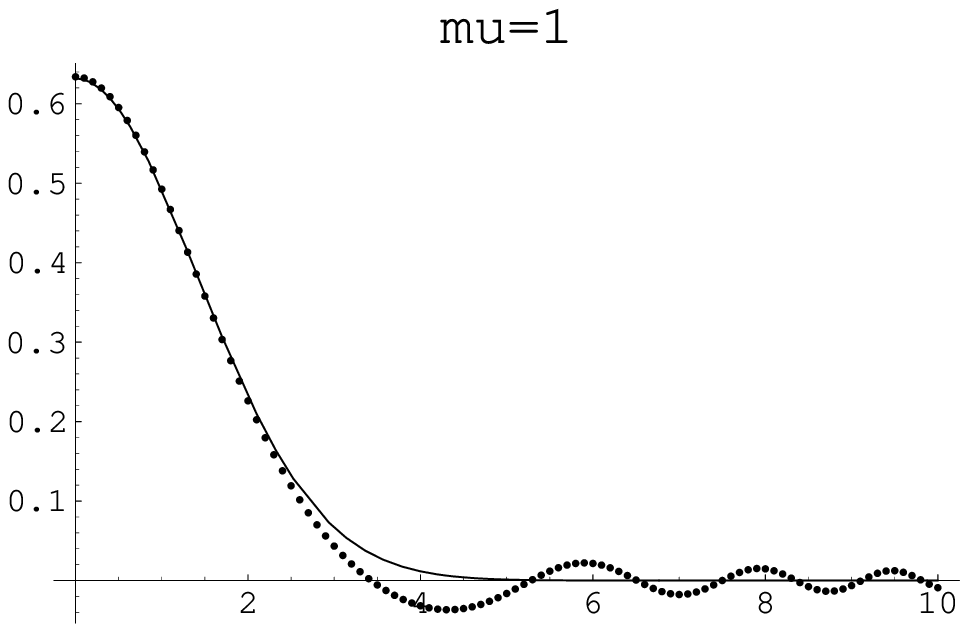}\hfill}
\vspace{-2cm}
\centerline{%
\hspace{-2cm}
\epsfysize=6cm \epsfbox{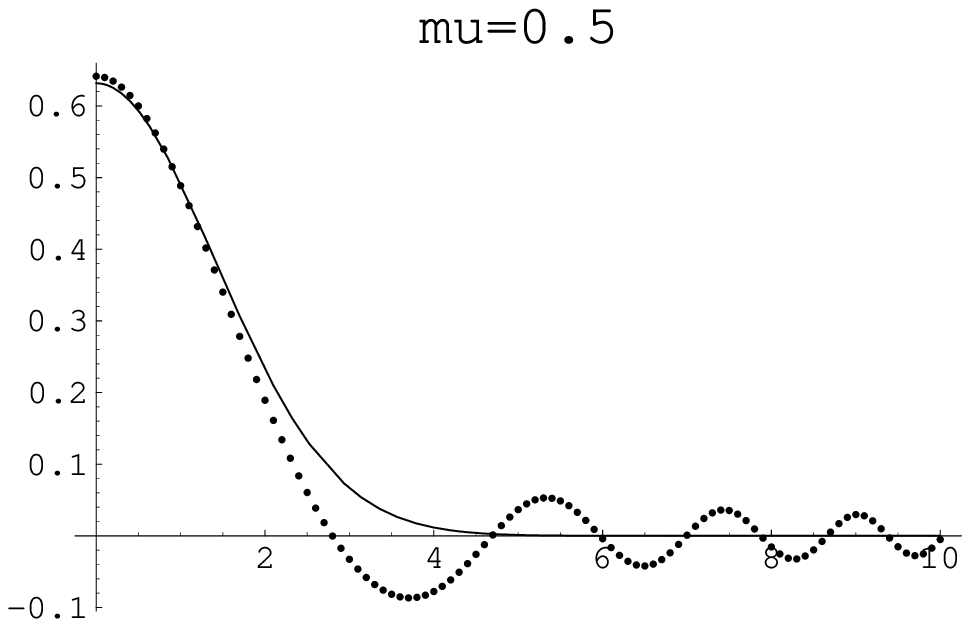}\hfill
\epsfysize=6cm \epsfbox{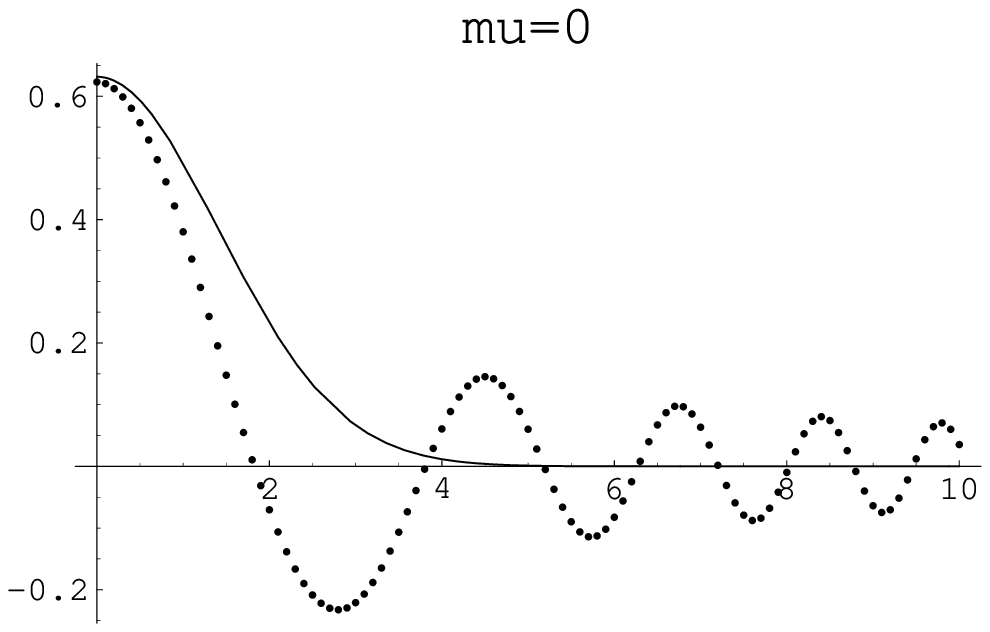}\hfill}
\caption{The wave-functions $\psi_{eff}(\lm,t=0)$ (dotted points)
  corresponding to the Gaussian wave-packet $\psi_0(\lm,t=0)$ 
in eq.\ \rf{psi0} (shown as a
  solid line) calculated for $\mu=1.5,1.0,0.5$ and $0$.}
\label{fig2}
\end{figure}

In order to extract $\psi_{eff}(\lm,t)$ from $\psi_0(\lm,t)$ one can
use the general formula \rf{int-rep} to get after some
manipulations\footnote{Write \rf{int-rep} as $\int_\mu^\infty d\mu
  \f{d}{d\mu} \psi_{proj}$, interchange the $\mu$ and $s$ integrations
  and change variables from $s$ to $v$.}:
\eq
\psi_{proj}(x,t)=\psi(x,t)- \f{i}{\sqrt{2} \pi^{\f{5}{4}}} e^{-\f{\pi}{2}\mu}
e^{i\mu t} \int_{-1}^1 \f{(1-v)^{-\f{3}{4}+i\mu}
  (1+v)^{-\f{3}{4}-i\mu} e^{i\f{x^2}{4} v}}{t+i\f{\pi}{2} +\log
  \f{1-v}{1+v}} dv
\eqx
Alternatively in this particular case one can explicitly find the 
decomposition in energy eigenstates\footnote{The parity 
even parabolic cylinder functions $\psi_E(\lm)$
can be found  in \cite{barton,moore}, we use here the normalization 
from \cite{moore} and $E=-\f{a}{2}$ where $a$ is the parameter used in
\cite{moore}.}:
\eq\label{psia}
\psi_0(\lm,t)= \int dE \, c_E \psi_E(\lm)e^{-iEt},~~~~~~~~~
c_E = 2\f{\sqrt{ \left| \f{\Gm(1/4-iE)}{\Gm(3/4-iE)} \right|}}{
 (4\pi \cosh(2 \pi E))^{\f{1}{4}}}
\eqx
where the coefficients fall off exponentially for 
large $|E|$, $c_E \sim |E|^{-{1}/{4}} e^{-{\pi |E|}/{2}}$.
For large $\mu$ the overlap with the Fermi sea will indeed be 
exponentially small in $\mu$ and thus ``non-perturbative'' in 
nature. However, the ``non-perturbative'' nature 
is of the same origin as most other ``non-perturbative'' 
corrections discussed in the literature since the exponential nature
of the correction comes from the Fermi-sea wave functions 
being in a classically forbidden region (around $\lm = 0$). 
The same can be said about the normalization constant $\NN$ from
\rf{NN}. Note also that the energy of this ``quantum'' brane will 
change from zero to  a positive value due to the interaction with 
the Fermi sea. 

In Fig.\ \ref{fig1} we have shown the behavior of 
the normalization constant $\NN$ and the energy $E_{eff}$ defined by
\eq
E_{eff} = \la \psi_{eff}| H |\psi_{eff}\ra.
\eqx
Only for $\sqrt{\al'}\mu < 1 $, i.e. for $g_s >1$ is there an effect which is 
not exponentially suppressed in $\mu$. This is corroborated by looking at the 
wave function $\psi_{eff}(\lm)$ itself. However, for small $\mu$ we see 
quite a large change in the wave function due to the interaction 
with the Fermi sea, as shown in Fig.\ \ref{fig2}.

\section{Discussion}
This work was motivated by the calculations of closed string
emission in the time-dependent rolling tachyon open string background
of Sen. These calculations, in the context of 26-dimensional bosonic 
string theory  were performed in \cite{llm} and in the context of 2d
critical string theory and/or matrix models in
\cite{verlinde,kms,many,tt}. 
The observation
was that the emitted energy was generically infinite. In the {\em concrete}
calculations the open string background was viewed as purely
classical\footnote{In \cite{kms} there is a discussion of using
  bosonization of chiral fermions to get a more complete treatment.}, 
which in the matrix model formulation 
translates into the statement that the eigenvalue corresponding to 
the D0-brane follows a classical trajectory in the inverted 
harmonic potential. Indeed, we saw that it was 
very simple to perform the closed string 
emission rate calculation using this classical trajectory.
Here again the emitted energy is infinite, in accordance
with the fact that the D0-brane is treated as a classical 
background object.
 
Since the {\it quantum states} of a D0-brane 
are in a one to one correspondence with the quantum states 
of the cut-off Hamiltonian in the inverted harmonic potential,
the matrix model offers a simple way to move away
from the classical situation. 
In this paper we derived an exact quantum description of the
time-dependent decay process of the unstable D0-brane in type 0B
two-dimensional superstring theory.
By doing so  
the emitted energy indeed becomes finite, and for small
$g_s$ it is cut off at $\sqrt{\al'} E = 1/g_s$. For large 
$g_s$ the cut-off is of order 1. 

As often before, we are 
in a situation where the matrix model offers us a simple 
way of addressing questions which are not 
easily addressed in the continuum or higher dimensional theory. 
In particular here we could exactly treat a quantum open string
background since, as discussed above, the quantum mechanics of
fermions in the upside-down harmonic potential with energy levels below 
the Fermi level at $E_F=-\mu$ filled
should be viewed as a candidate for the continuum 
quantum open string field theory
in 2d, describing the dynamics of D0-branes. 

In order to apply the lesson from matrix models to higher dimensional critical 
string theories one needs to understand how to describe
the quantum D-brane (quantum open string background) from a string
point of view. One possibility is to 
use Witten's open string field theory at the quantum level. However it
would be most attractive to have a direct {\em worldsheet} description.
Presently we can understand that the classical D-brane
should be associated with a boundary conformal field theory,
but the quantum D-brane generalizes this situation, and there should 
be a suitable string-theoretical description of the quantum 
D-brane. The ease with which the situation is handled 
in the matrix model context and the fact that very 
sensible results emerge is a strong hint that there should 
exist a simple higher dimensional string description too.   

\subsection*{Acknowledgment}
Both J.A.\ and R.J.\  acknowledge support by the
EU network on ``Discrete Random Geometry'', grant HPRN-CT-1999-00161
as well as  by {\it MaPhySto}, Network of Mathematical Physics 
and Stochastics, funded by a grant from Danish National Research Foundation.
R.J. was supported in part by  KBN grants~2P03B09622 (2002-2004),
2P03B08225 (2003-2006).

\end{document}